# Efficient Neighbor Discovery for Proximity-Aware Networks

*C. Jiang, L. Yang*

*Abstract*—**In this work, we propose a fast and energy-efficient neighbor discovery scheme for proximity-aware networks such as wireless ad hoc networks. Discovery efficiency is accomplished by the use of a special discovery signal that provides random multiple access with low transmit power consumption and low synchronization requirement.**

## I. Introduction

Knowledge of neighbors is essential for all proximity-aware networking, such as wireless ad hoc sensor networks and social networks. In ad hoc networks, awareness of neighboring devices/nodes is the key to routing protocols, medium access control protocols and some topology-control algorithms. The device can be a sensor in a sensor network, a user's smart phone in a social network, a home equipment in a home network, or a vehicle in a vehicular network. Neighbor discovery, i.e., the detection of a device's immediate neighbors, is therefore a crucial first step in the process of self-organization of a wireless ad hoc network [1]. Due to the mobile nature of a wireless ad hoc network, such as the vehicular network and the social network, the topology of the network constantly changes. Even for many sensor networks with the static nature, connectivity is still subject to changes even after the network has been established [2]. The devices/nodes must look for neighbors on a regular basis to accommodate network topology changes and re-synchronize to neighbors due to accumulated clock drifts.

Energy conservation is critical for almost all wireless devices. This is especially true for the case of wireless sensor networks where sensor nodes must operate using finite, non-replaceable energy sources.

Energy efficiency ultimately determines the lifetime of a sensor network and thus has top priority when designing wireless sensor networks. Since neighbor discovery is an on-going, periodical operation, energy efficiency of the discovery process thus has a large effect on the lifetime of the device.

In this paper, we propose an energy-efficient neighbor discovery scheme for use in proximity-aware networking. The rest of this paper is organized as follows. Section II gives a brief description of the basic neighbor discovery scheme providing a baseline performance for neighbor discovery. Section III describes the proposed efficient neighbor discovery scheme. Section IV provides a numerical example and simulation results to validate the proposed scheme. Section V concludes this paper.

## II. BASIC NEIGHBOR DISCOVERY SCHEME

This section provides a brief review of the framework that constitutes the foundation for neighbor discovery of most existing ad-hoc networks [1]-[5].

The goal of the neighbor discovery process is to have each mobile device in the network discover all of its immediate neighbors (the nodes with which it can establish a direct wireless communication). Assume all nodes in the neighborhood follow a common slotted channel structure as illustrated in Fig. 1. One of every other $T$ slots is designated for neighbor discovery. $T$ is typically large to ensure low duty cycle. Devices who are only interested in discovery can go to sleep between discovery slots for power saving. In a typical neighbor discovery scheme, a device transmits its discovery signal in the discovery slot with probability $p$ and listens for transmissions from other devices with probability $1-p$. Collisions occur if a device simultaneously receives transmissions from two or more of its neighbors. Under the assumption that transmissions of discovery signals among devices are independent, the probability that device $i$ discovers its neighbor device $j$ out of $L$ total neighboring devices within $t$ discovery slots is given by

$$P_{ij} = 1 - \left(1 - p(1-p)^{L-1}\right)^t, \quad \forall i,j \qquad (1)$$

The upper bound of the above probability,

$$\max_{p} P_{ij} = 1 - \left(1 - L^{-1}\left(1 - L^{-1}\right)^{L-1}\right)^{t}, \qquad (2)$$

becomes small for large *L*. Hence, the discovery time of a device can be very long. Large *L* also causes more collisions, and, therefore, many wasteful transmissions of discovery signals as a result of collisions.

It is clear that collision is very expensive in terms of energy consumption and thus should be kept minimum, which is indeed one of the main objectives for the design of the proposed neighbor discovery scheme.

## III. THE PROPOSED NEIGHBOR DISCOVERY SCHEME

A. TNID and Discovery Channels

As a first step to avoid the collision, a temporary node ID (TNID) is created for identifying the nodes in a neighborhood and is used for all communications among devices. The same TNID can be spatially reused in a different neighborhood. In the proposed scheme, each discovery slot is divided into *D* parallel channels, each of which uniquely represents a TNID. A node broadcasts its presence and its associated TNID by energizing one of the TNID channels.

To create the *D* parallel channels for discovery signaling, the discovery slot is first divided into multiple OFDM symbols. A large fraction, if not all, of the energy in an OFDM symbol is transmitted on a single OFDM subcarrier. No energy is transmitted on any other subcarriers of the current OFDM symbol. No information is modulated onto the energized subcarrier (i.e. amplitude and/or phase), therefore no channel estimation is needed at the receiver and low dynamic range and phase noise requirements for the PA (power amplifier) and RF (radio frequency) components. It is the location (subcarrier index) of the energized tone that contains information. That is, which subcarrier of this OFDM symbol is energized depends on the content of the message. In the current application, the message contains TNID and is denoted as *m*, which is further represented by *K* information symbols, $\mathbf{u} = (u_1, u_2, ..., u_K)$ for $1 \leq k \leq K$, or more precisely,

$$m(D) = u_K D^{K-1} + u_{K-1} D^{K-2} + ... + u_2 D + u_1 \tag{3}$$

where the base *D* is the total number of subcarriers used for transmitting one information symbol $u_k$ ($0 \leq u_k \leq D-1$). For the ease of discussion, we assume that the total number of subcarriers of an OFDM symbol *S* is also equal to *D*. We thus need *K* OFDM symbols to transmit the message *m*, as shown in Fig. 2. Note that this single-tone signal is not to be confused with a frequency hopping signal. In frequency hopping, the tone positions are predetermined by a sequence known to both the transmitter and the receiver; the tone position therefore does not contain information and the information is modulated onto the amplitude and/or phase of the tone via QPSK or QAM. While for the single-tone scheme, the tone is not modulated with information. The information is embedded in the positions/subcarrier indices of the tones.

The choice of this type of signal has the following advantages. First of all, unlike the most commonly used CDMA signals for random access [6]-[14], the single-tone signal does not suffer from the near-and-far effect among transmissions from different users. This is because 1) If the single-tone signals from different users are transmitted on different subcarriers, they don't present interference to each other since all the OFDM subcarriers are orthogonal (Fig. 3); 2) If some of the tones from different users happen to transmit on the same subcarrier, they simply add together just like multi-paths (since the tones are not modulated therefore share the same waveform and are not distinctive at the receiver) and are absorbed by cyclic prefix [15]. Second, the use of the cyclic prefix of an OFDM symbol, relaxes the time synchronization requirement among devices. The longer the cyclic prefix, the more tolerance to timing offset among devices, therefore, the less power spent on time synchronization and less receiver complexity and energy consumption. Third, this single-tone (narrow band) waveform has the lowest peak-to-average power ratio (PAPR) as compared to a wideband waveform (e.g., the regular OFDM waveform) allowing for most efficient power usage;

However, single tone signaling can cause ambiguity among different discovery signals from multiple devices at the receiver. As illustrated in Fig. 3, it is difficult for the receiver to figure out which set of tones comprises the original sequence sent by a device since the individual tones themselves does not identifies the device (although in the figure different devices are identified by color). The possible combinations are $t^K$ where $t$ is the number of simultaneous transmitting devices ($t = 2$ in Fig. 3). In addition, single tone signaling can also be prone to errors due to noise and fading causing missing tones and/or falsely detected tones. The solution proposed in this paper is to encode the discovery signal (non-binary) to prevent the ambiguity and to obtain certain degrees of error protection capability.

Reed-Solomon codes are non-binary codes and achieve the largest possible code minimum distance for any linear code with the same encoder input and output block lengths (or maximum distance separable). Hence the Reed-Solomon code is a good fit for the current application. A Reed-Solomon code $(N, K)$ encodes the non-binary information symbols $\mathbf{u} = [u_1, u_2, ..., u_K]^T$ into a codeword $\mathbf{c} = [c_1, c_2, ..., c_N]^T$, where $0 \leq c_n \leq D-1$ for $1 \leq n \leq N$. The error-correction capability is $t = \left\lfloor \dfrac{N-K}{2} \right\rfloor$ and the erasure-correction capability is $\rho = N - K$, where $\lfloor x \rfloor$ denotes the maximum integer which does not exceed $x$.

Let $\alpha$ be a primitive number in GF($D$), the $K$ information symbols $\mathbf{u} = (u_1, u_2, ..., u_K)$ are encoded into an $N$ coded symbols via the Galois Fourier Transform

$$\mathbf{c}^T = \mathbf{Z} \begin{bmatrix} 0 & \mathbf{u}^T & 0 & \cdots & 0 \end{bmatrix}^T \qquad (4)$$

where

$$\mathbf{Z} = \begin{bmatrix} 1 & 1 & \cdots & 1 \\ 1 & \alpha^{\frac{D-1}{N}} & & \alpha^{\frac{D-1}{N}(N-1)} \\ \vdots & & \ddots & \vdots \\ 1 & \alpha^{\frac{D-1}{N}(N-1)} & \cdots & \alpha^{\frac{D-1}{N}(N-1)(N-1)} \end{bmatrix} \quad (5)$$

The value of the code symbol ($c_n$, $1 \leq n \leq N$) corresponds to the index of the sub-carrier on which energy is transmitted. That is, only one tone is energized per OFDM symbol, and the position of the tone is dependent on the value of the code symbol. Either all or partial of the total energy in an OFDM symbol is transmitted on a single subcarrier depending on the desired coverage range of the neighborhood. Fig. 4 shows an example of the transmission of the discovery signal. The detection of a discovery signal tone is done by simply looking for a subcarrier with significantly higher energy than its neighbors. After the detection of discovery tones, the receiver obtains a set of discovery tones on every OFDM symbol with certain errors as well as missed discovery tones. By applying, for example, maximum likelihood decoding using a lookup table, the receiver finally recovers the original information symbols and therefore the TNID.

In the presence of a plural discovery signals, the discovery signals from different nodes may overlay on top of each other causing potential interference among different discovery channels. Indeed, $d$ discovery signals with code rate ($N$, $K$) can coexist without causing decoding ambiguity as long as the following inequality

$$K \leq \left\lceil \frac{N}{d} \right\rceil \quad (6)$$

is satisfied. This important conclusion can be formally stated by the following proposition:

*Proposition: Assume $d$ ($d \leq D^K$) distinctive discovery signals coded on GF(D) with code rate $(N, K)$ are simultaneously received on the same time and frequency resource. Under perfect tone detection, all d coded discovery signals can be decoded to the original information symbols without ambiguity, if*

$K \leq \left\lceil \dfrac{N}{d} \right\rceil$ *is satisfied.*

*Proof:* Consider $d$ $(d \leq D^K)$ distinctive discovery signals coded with rate $(N, K)$ on *GF(D)* are simultaneously received from *N* OFDM symbols, free of tone erasures and detection errors. Now arbitrarily select *N* number of the detected tones, each from one of the *N* different OFDM symbols. We maintain that

1) There are at least $\left\lceil \dfrac{N}{d} \right\rceil$ discovery tones out of the *N* selected tones coming from the same discovery signal among the total number of *d* discovery signals. This is the direct outcome from the pigeonhole principle.

2) For an discovery signal with code rate of $(N, K)$, a minimum number of *K* discovery tones is sufficient to distinguish one discovery signal from another. This is sustained by the fact that Reed-Solomon codes are maximum distance separable,

We therefore conclude that if $\left\lceil \dfrac{N}{d} \right\rceil \geq K$, the *d* discovery signals can be uniquely separated from each other without ambiguity. □

For example, $(16, 2)$ coded discovery signals can allow up to 15 simultaneous discovery signal transmissions from different nodes, i.e., providing 15 orthogonal discovery channels.

However, the case of particular interest is *K*=1. When *K*=1, (6) holds for any value of *d* regardless the value of *N*. We therefore have the following important remark:

*Remark 1*: *Assume d ($d \leq D$) distinctive discovery signals coded on GF(D) with code rate $(N,1), \forall N \geq 1$ are simultaneously received on the same time and frequency resource. Under perfect tone detection, all d discovery signals can be decoded to the original information symbols without ambiguity.*

There are total of *D* discovery channels representing *D* TNIDs. The *D* discovery signals therefore constitute *D* discovery channels. Two nodes transmitting at different discovery channels (therefore different TNIDs) do not collide with each other. For example, (8, 1) coded discovery signals in theory can support up to $D$ number of simultaneous discovery signal transmissions, or, $D$ discovery channels. This is true whether the value of $N$ equals to 8 or not.

However, this conclusion is only true under the assumption of ideal tone detection. In practical scenarios *when tone detection is not error-free* due to fading and noise, the value of $N$ *does* affect the discovery signal's capability of correcting tone detection errors. The value of $N$ and the capability of tone error correction and erasure recovery have been discussed earlier in this section. Therefore, as long as the number of tone detection errors/erasures are within the capability of the coded discovery signals governed by the value of $N$, multi-user ambiguity can be eliminated.

Another important issue that can not be overlooked in the wireless ad hoc network signaling design is the time and frequency synchronization. Since the discovery signal is meant to be received by neighboring devices, the discovery signal has to be designed with time and frequency offset tolerance. The time offset among devices is easily absorbed by the cyclic prefix of the OFDM symbol as earlier stated thereby is less of a concern. The design of STS with frequency offset immunity is more challenging since the information that an STS signal carries lies in the subcarrier index of the STS tones. If not appropriately designed, the STS tones may shift to neighboring subcarriers as a result of frequency offset. Again, we have to rely on coding to provide the frequency offset immunity.

Assume that the transmitted discovery signal is

$$\mathbf{c} = \begin{bmatrix} c_0 & c_1 & \cdots & c_{N-1} \end{bmatrix}. \tag{7}$$

The received discovery signal at the node with frequency offset $\delta$ is

$$\mathbf{c}' = \begin{bmatrix} c_0 + \delta & c_1 + \delta & \cdots & c_{N-1} + \delta \end{bmatrix} \tag{8}$$

From (4), the inverse GFT of (7) is given by

$$\mathbf{Z}^{-1}\mathbf{c}^T = \begin{bmatrix} 0 & \mathbf{u}^T & 0 & \cdots & 0 \end{bmatrix}^T \tag{9}$$

with the first element equal to zero. Therefore, the inverse GFT of a valid code word always has a zero-valued first element *by design*. Whereas the first element of the inverse GFT of (8) produces

$$\frac{1}{N}\sum_{n=0}^{N-1} c'_n = \frac{1}{N}\sum_{n=0}^{N-1}(c_n + \delta) = \frac{1}{N}\sum_{n=0}^{N-1} c_n + \delta = \delta \tag{10}$$

which is non-zero. We therefore have the following remarks:

*Remark 2: A discovery signal received with a frequency offset does not correspond to a valid but wrong discovery signal.*

This property ensures that a receiver with frequency offset will not erroneously map a discovery signal to a valid TNID.

*Remark 3: The value of the first element of the inverse GFT of a received discovery signal equals to the frequency offset between the transmit node and the receive node.*

This property enables the receiver to correct the frequency offset, if any, between the transmit node and the receive node. The frequency-offset discovery signal can then be recovered. The property is hence very beneficial to initial neighbor discovery when a node is not frequency-synchronized to its neighbors.

B. TNID Acquisition and Hidden Device Avoidance

When entering the network, a device first synchronizes to the slot by synchronizing to the neighbors' discovery signals. To transmit its own discovery signal, a device must first acquire a vacant TNID (or a vacant discovery channel) from the pool of $D$ valid TNIDs.

During acquisition, device $i$ scans through the $D$ discovery channels searching for the one with the lowest average energy over a certain number of discovery slots (i.e., the least congested TNID), i.e.,

$$d^* = \arg\min_{d \in \{0,1,\cdots,D-1\}} \sum_j \sum_{n=0}^{N-1} \left| h_{ji}\left(c_n^d\right) + w \right|^2, \tag{11}$$

where $c_n^d$ is the *d*th discovery signal tone index, $h_{ji}(c_n^d)$ is the channel gain at subcarrier $c_n^d$ between node *i* and node *j*, and *w* is the receive noise spectral power density. To reduce the probability that two or more devices simultaneously acquire the same TNID with the lowest energy, the device randomly selects a TNID from a set of channels with the lowest energies. Once a TNID is acquired, the device starts to transmit the corresponding discovery signals with probability *p*.

It is still possible that two or more devices simultaneously select the same TNID. A device therefore listens for collision by measuring energy on this discovery channel on the discovery slots. A device makes a decision based on the measured energy if a collision has happened and if a collision is significant enough (the receive energy from the colliding node has significant strength) to warrant a re-acquisition of a new TNID.

Since TNIDs can be spatially reused, hidden device problem may occur. As illustrated in Fig. 5, device A and device C may acquire the same TNID since device C cannot hear device A and vice versa. However, in this particular device distribution, device B can hear both, thereby causing collision at device B, i.e., the same TNID is used by two *immediate* neighbors of node B. To prevent the hidden device problem, a device (e.g., device B) listens on other discovery channels to measure the transmission probability of the discovery signal on each channel. If the probability is much larger than *p*, two or more devices may have already been using the same TNID without being aware of the collision at another device (i.e., device B). The device (device B) thus jams this discovery channel by transmitting a discovery signal corresponding to the collided TNID. The hidden devices (device A and C) detect the jammer, realizing a TNID collision, and consequently restart the TNID acquisition procedure.

## IV. NUMERICAL RESULTS

As earlier stated, the message of the discovery signal contains the TNID. In the simulation, a total of

512 TNID numbers were used ranging from 0 to 511 and coded with code rate $(8, 1)$. The total number of subcarriers is 512. One discovery tone (i.e., one discovery code symbol) can then be transmitted on one OFDM symbol.

Fig. 6 shows the performance of the proposed discovery signal, in which the number of simultaneously transmitting devices is 30 and the number of receive antennas per device is one. Performance in AWGN channel is plotted as a reference. SNR is defined as the ratio of received energy per sample to noise variance in time domain. A decoding erasure is defined as the event in which a device fails to decode the discovery signal sent from another device, while a decoding error is an event in which the device decodes the discovery signal to a wrong but valid TNID. An erasure prevents the device from recognizing the neighbor whereas an error causes the device to see a neighboring node that does not exist. In Fig. 6, the error rate is controlled below 1%. It is observed that the discovery signal can operate at very low SNR. In the low SNR region, the up-fades from the frequency selective fading in PedB channel create more opportunities than AWGN channel for the discovery tones to be detected, causing less erasure. Fig. 7 plots the median of neighbor discovery time vs. the density of the device drops. The devices are uniformly dropped over an area of 1024 with various densities. The superior advantage of the proposed scheme is clearly seen over the conventional baseline method (Section II). This is not totally surprising due to the fact that multiple discovery signals can be transmitted simultaneously without collision and hence multiple devices can be discovered in a single discovery slot.

## V. CONCLUSIONS

In this paper, we proposed a neighbor discovery scheme for proximity-aware networks. This scheme allows simple transceiver implementation and most importantly energy efficiency. The energy efficiency are achieved by the use of a specially designed discovery signal. In particular, the energy efficiency is achieved by: 1) Using energy efficient discovery signals; 2) Reducing the number of wasteful transmissions of discovery signals. The energy efficient discovery signal is obtained by 1) using low-PAPR single-tone pulse. 2) Unlike the conventional signaling, the amplitude and phase of the tone (susceptible to interference) is not used for conveying information. The information is instead modulated into the position of the tone that reduces the transceiver complexity and increases immunity to noise allowing to be detected at low signal to noise ratio. The resulting ambiguity at the receiver is prevented by the use of non-binary coding. 3) Discovery signal is robust to frequency and time offset that relaxes the requirement for synchronization. Signal transmission is the most energy-consuming operation thereby should be minimized. The reduction in wasteful discovery signal transmissions without causing excessive discovery delay is due to the collision prevention among discovery signals that is provided by 1) the use of TNID and its associated discovery channels, requiring a device to acquire a vacant TNID first before it can transmit the discovery signal; 2) Multiple discovery signals can be transmitted simultaneously within the same discovery slot without collision. Finally, energy efficiency is also attributed to discovery efficiency, a result from the fact that multiple devices can be discovered during a single discovery slot. Therefore a device can spend less time on discovery (receive mode).

## REFERENCES

[1] S. Vasudevan, J. Kurose, and D. Towsley, "On neighbor discovery in wireless networks with directional antennas," *Proc. IEEE INFOCOM*, 2005, vol. 4, pp. 2502-2512.

[2] R. Cohen and B. Kapchits, "Continuous neighbor discovery in asynchronous sensor networks," *IEEE Trans. Networking*, pp. 69-79, no. 1, Feb. 2011.

[3] R. Madan and S. Lall, "An energy-optimal algorithm for neighbor discovery in wireless sensor networks," *Mobile Network Applications*, vol. 11, no 3, pp. 317-326, 2006.

[4] A. Keshavarzian, E. Uysal-Biyikoglu, f. Herrmann, and A. Manjeshwar,"Energy-efficient link assessment in wireless sensor networks," *Proc. IEEE INFOCOM*, 2004, vol. 3, pp. 1751-1761.

[5] E. Hamida, G. Chelius, and E. Fleury, "Revisiting neighbor discovery with interferences consideration," *Proc. PE-WASUN*, 2006, pp. 74-81.

[6] P. Zhou, H. Hu, H. Wang, and H.-H. Chen, "An efficient random access scheme for OFDMA systems with implicit message transmission," *IEEE Trans. Wireless Commun.*, vol. 7, no. 7, pp. 2790-2797, July 2008.

[7] Y. Yang and T.-S. P. Yum, "Analysis of power ramping schemes for UTRA-FDD random access channel," *IEEE Trans. Wireless Commun.*, vol. 4, no. 6, pp. 2688- 2693, Nov. 2005.

[8] C.-S. Hwang, K. Seong, and J. Cioffi, "Throughput maximization by utilizing multi-user diversity in slow-fading random access channels," *IEEE Trans. Wireless Commun.*, vol. 7, no. 7, pp. 2526-2535, July 2008.

[9] F. Khan, *LTE for 4G Mobile Broadband: Air Interface Technologies and Performance*, Cambridge University Press, 2009.

[10] V. Aggarwal and A. Sabharwal, "Performance of multiple access channels with asymmetric feedback," *IEEE J. Sel. Areas Commun.*, vol. 26, no. 8, pp. 1516-1525, Oct. 2008.

[11] J.-P. M. G. Linnartz, R. Hekmat, and R.-J. Venema, "Near-far effects in land mobile random access networks with narrow-band Rayleigh fading channels," *IEEE Trans. Veh. Technol.*, vol. 41, no. 1, pp. 77-90, Feb. 1992.

[12] H. H. Chen, *the Next Generation CDMA Technologies*, John Wiley & Sons, 2007.

[13] H. Holma, A. Toskala, *WCDMA for UMTS: HSPA Evolution and LTE*, John Wiley & Sons , 2010.

[14] A. Ghosh, J. Zhang, J. Andrews, R. Muhamed, *Fundamentals of LTE*, Prentice Hall, 2010.

[15] D. Tse and P. Viswanath, *Fundamentals of Wireless Communications*, Cambridge University Press, 2005.

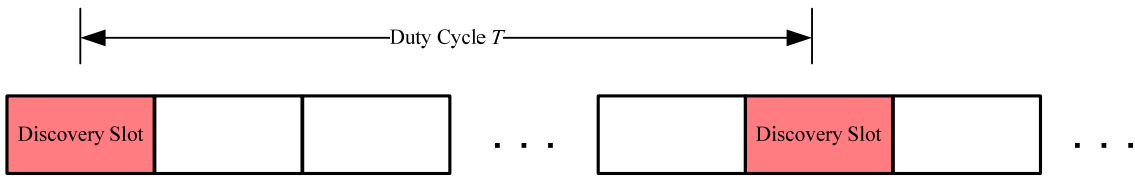

Fig. 1. Illustration of discovery slots.

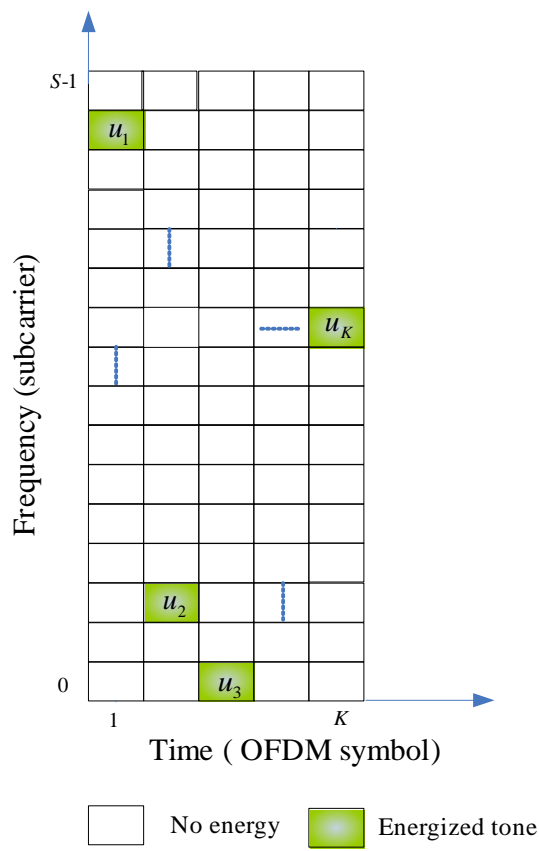

Fig. 2 Illustration of a neighbor discovery signal, where $u_k$, $1 \leq k \leq K$, is the *index* of the subcarrier that the discovery signal tone is transmitted on.

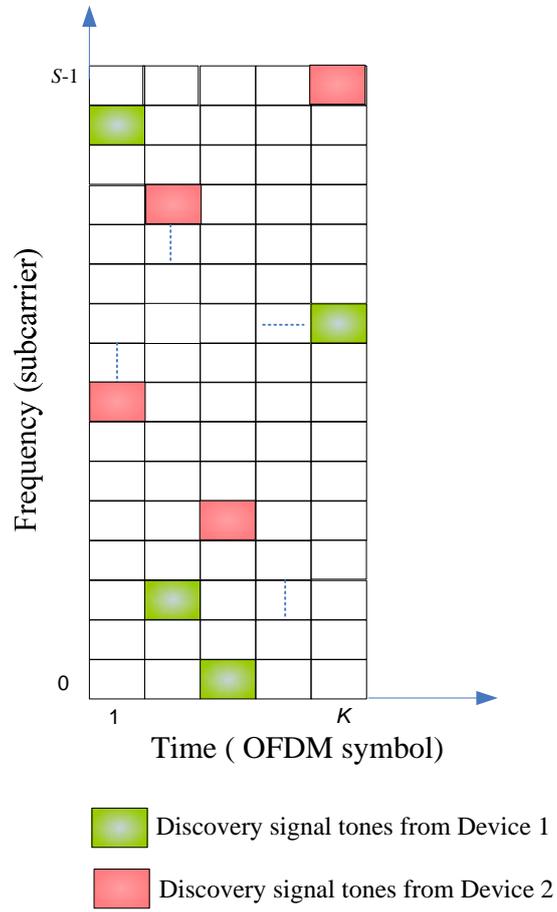

**Fig. 3. Illustration of discovery signal from two devices.**

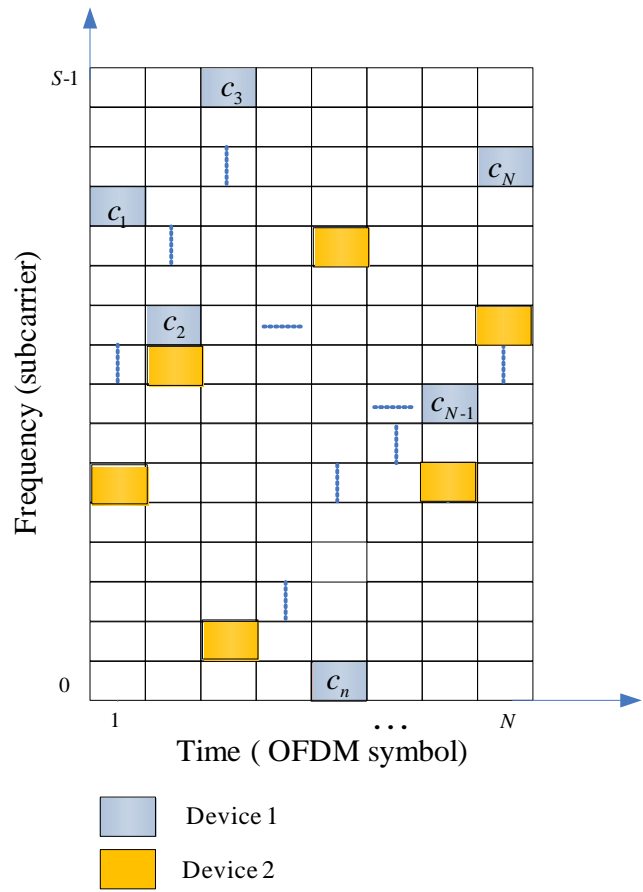

**Fig. 4. Illustration of coded discovery signals at a receiver, where** $c_n$, $1 \leq n \leq N$, **is the *index* of the subcarrier that the discovery signal tone is transmitted on.**

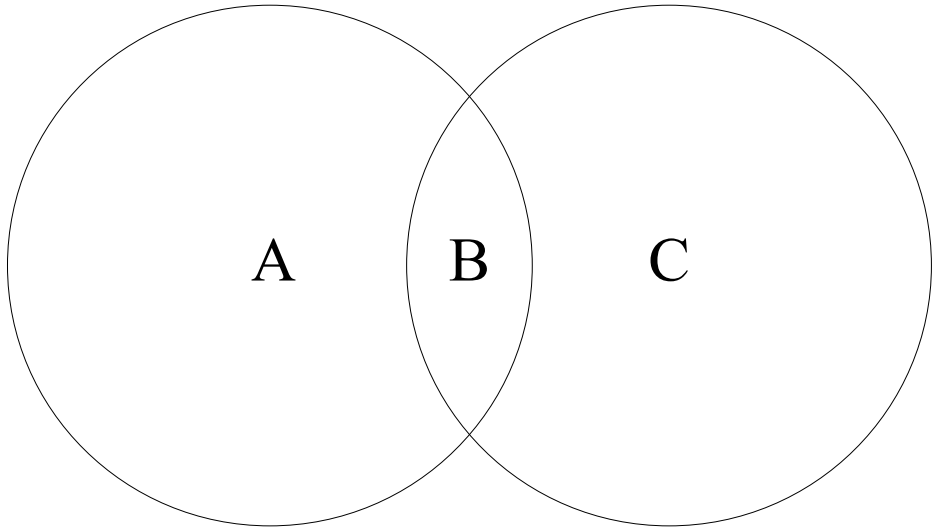

**Fig. 5. Illustration of hidden devices.**

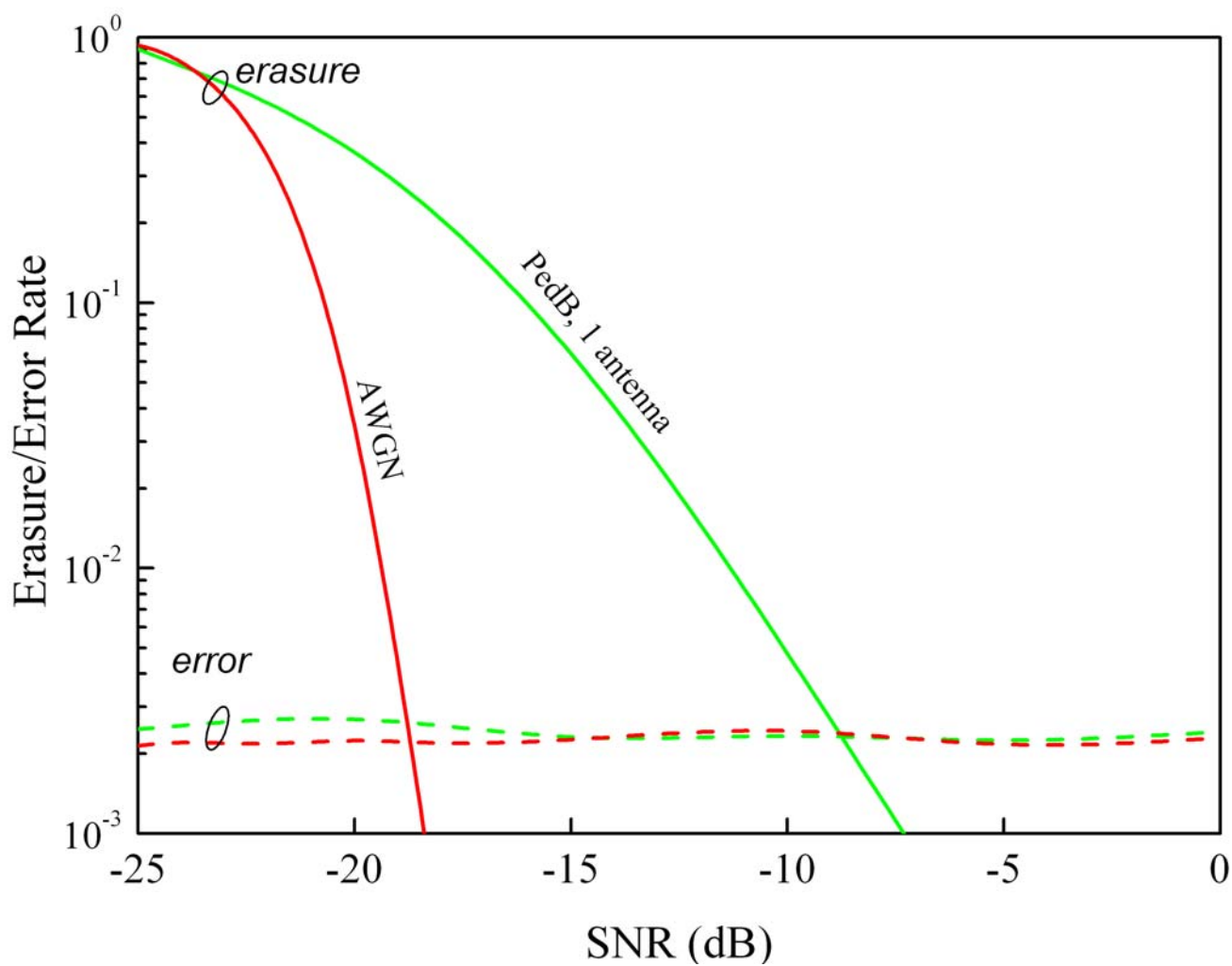

Fig. 6. The decoding erasure and error performance of the coded discovery signaling in a multi-node scenario (30 nodes; total TNIDs = 512; number of subcarriers in an OFDM symbol = 512, code rate of discovery signal $=(8,1)$, fading speed = 3 km/h at 2 GHz carrier frequency; number of receive antennas per device =1).

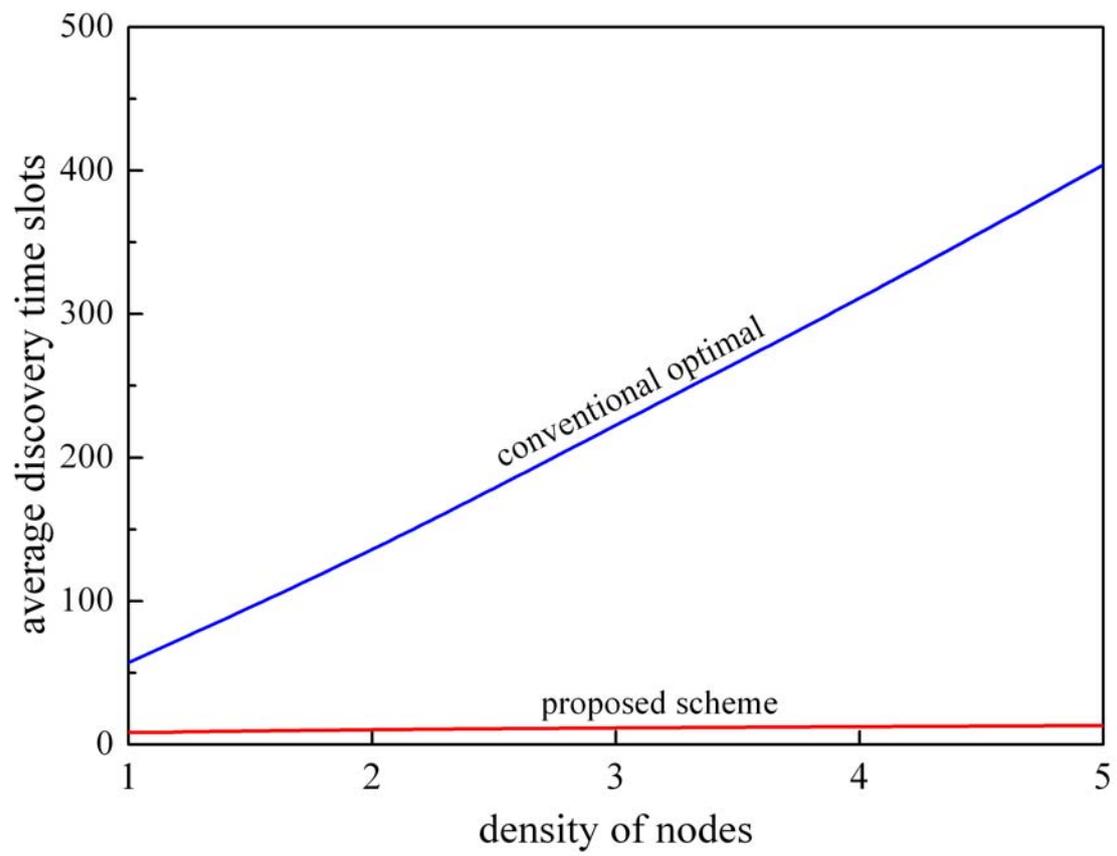

**Fig. 7. Discovery delay performance vs the density of device/nodes.**